# On the origin of the hydraulic jump in a thin liquid film

R. K. Bhagat[1*], N. K. Jha[2], P. F. Linden[2] and D. I. Wilson[1]

[1] Department of Chemical Engineering & Biotechnology, University of Cambridge, West Cambridge site, Philippa Fawcett Drive, Cambridge, CB3 0AS

[2] Department of Applied Mathematics and Theoretical Physics, Centre for Mathematical Sciences, Wilberforce Road, Cambridge CB3 0WA

**Abstract**

For more than a century, it has been believed that all hydraulic jumps are created due to gravity. However, we found that thin-film hydraulic jumps are not induced by gravity. This study explores the *initiation* of thin-film hydraulic jumps. For circular jumps produced by the normal impingement of a jet onto a solid surface, we found that the jump is formed when surface tension and viscous forces balance the momentum in the film and gravity plays no significant role. Experiments show no dependence on the orientation of the surface and a scaling relation balancing viscous forces and surface tension collapses the experimental data. Experiments on thin film planar jumps in a channel also show that the predominant balance is with surface tension, although for the thickness of the films we studied gravity also played a role in the jump formation. A theoretical analysis shows that the downstream transport of surface tension energy is the previously neglected, critical ingredient in these flows and that capillary waves play the role of gravity waves in a traditional jump in demarcating the transition from the supercritical to subcritical flow associated with these jumps.

1. **Introduction**

When a jet of water falls vertically from a tap on to the base of a domestic sink, the water spreads radially outwards in a thin film until it reaches a radius where the film thickness increases abruptly. This abrupt change in depth is the *circular hydraulic jump*. A similar phenomenon is observed on vertical and inclined surfaces (including urinal walls), where the liquid film spreads radially outwards before forming a jump.

The hydraulic jump has been studied for over four hundred years. An early account was presented by Leonardo de Vinci in the 16th century[1]. The Italian mathematician Giovanni Giorgio Bidone (1819) [2] published experimental results on the topic and Lord Rayleigh (1914) subsequently provided the first theoretical explanation for the planar hydraulic jump based on inviscid theory[2,3].

All existing theories invoke gravity in the origin of the hydraulic jump[4,5] implying that the hydraulic jump location should be sensitive to the orientation of the surface. However, we observed that, under the same flow conditions, normal impingement of a liquid jet gives a circular hydraulic jump with the same initial radius irrespective of the orientation of the surface. On a vertical plate, where the spreading liquid film and gravity are coplanar, an approximately circular hydraulic jump is formed initially (Fig 1 (*a*)). The thick liquid film beyond the hydraulic jump then drains downwards due to gravity [6-9]. On a horizontal surface, the jump stays at the same location (Fig 1 (*b*)) until the liquid reaches the edge of the plate, which changes the downstream flow and the subsequent position of the jump. Similarly, when a jet impinges onto a horizontal surface from below, a circular hydraulic jump is formed (Figure 1(*c*)). Under the influence of gravity, the thick liquid film beyond the hydraulic jump falls as droplets or as a continuous film forming a water bell [10]. Figure 1 shows that in all three cases the hydraulic jump has almost the same radius ($R \approx 26$ mm). These three experiments show unequivocally that *gravity plays no role in the formation* of the circular hydraulic jump in a thin liquid film and that gravity only affects the jump *after* it is formed.

Watson (1964) proposed the first description of a thin-film circular hydraulic jump (such as in a sink),

*Correspondence to:* rkb29@cam.ac.uk

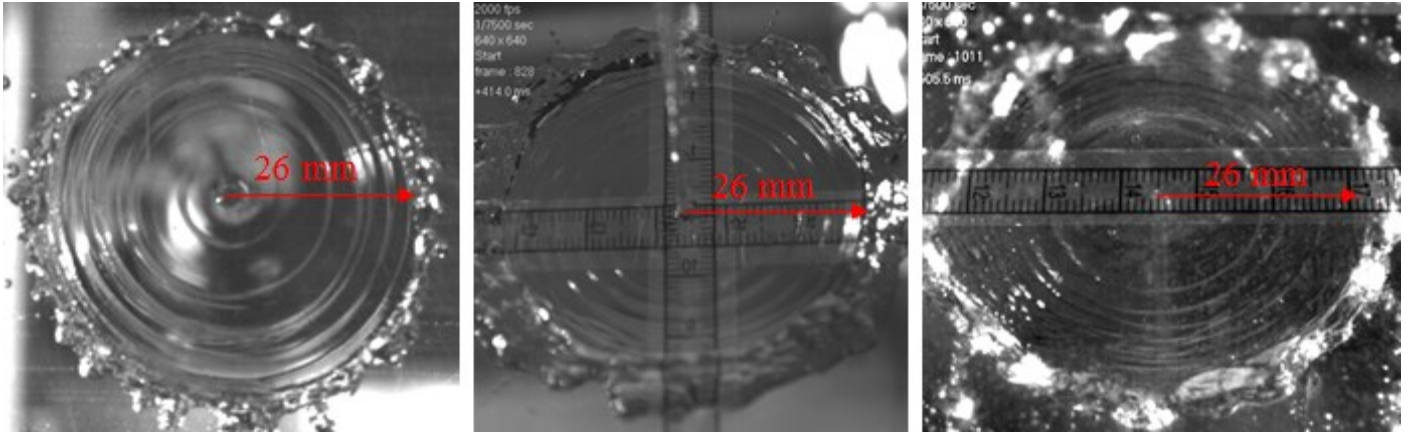

Figure 1. Hydraulic jumps caused by a water jet impinging normally to (*a*) a vertical surface, viewed from the side; (*b*) a horizontal surface, viewed from above; and (*c*) a horizontal surface, jet impinges from beneath, viewed from above. In these cases the jets are identical, produced from the same nozzle at the same flowrate *Q* = 1 l/min, and the radius of the jump is independent of the orientation of the surface.

incorporating the viscous friction in the thin liquid film and balancing the momentum and hydrostatic pressure across the jump [11]. Watson's solution, which involves gravity, cannot, however, predict the jump radius without experimental measurement of the film thickness at the jump location, and it overpredicts the jump radius for smaller jumps by as much as 50% [12]. Bush & Aristoff (2003) incorporated the effect of surface tension in Watson's theory but argued that its influence was small as its effect was confined to the hoop stress associated with the increase in circumference of the jump. Bohr *et.al.* (1993) connected the inner and the outer solutions for radial flow through a shock in shallow water and obtained a scaling relation $R \sim Q^{5/8} \nu^{-3/8} g^{-1/8}$ where $R, Q, \nu$ and $g$ are the jump radius, the jet volume flux, the kinematic viscosity fluid and gravitational acceleration, respectively[4]. They showed that the outer solution for the hydraulic jump becomes singular at a finite radius where the local Froude number, *Fr* = 1. They argued that the jump could be understood qualitatively in terms of the interplay between gravity and the momentum of the liquid.

The experimental observations presented in this paper show a sharp departure from these approaches. Furthermore, the existing theories require information or feedback from the liquid film downstream of the hydraulic jump to predict its location[4,5], but the initially spreading liquid film does not receive information of this nature. We present here a new scaling relation and a theoretical approach that explains the *initial location* of the jump and compare its predictions with experimental results obtained with liquids of different viscosity and surface tension.

## 2. Scaling analysis

Consider a cylindrical co-ordinate frame with *r* and *z* the radial and jet-axial coordinates, respectively, *u* and *w* the associated velocity components, and assume circular symmetry about the jet axis. In the boundary layer approximation, the equations governing flow in a thin film are

$$\frac{\partial (ru)}{\partial r} + \frac{\partial (rw)}{\partial z} = 0, \quad (1)$$

$$u\left(\frac{\partial u}{\partial r}\right) + w\left(\frac{\partial u}{\partial z}\right) = -\frac{dp}{dr} + \nu\left(\frac{\partial^2 u}{\partial z^2}\right) - g\frac{dh}{dr} \quad (2)$$

where $h(r)$ is the thickness of the film, and the gauge pressure *p* arises from the local film curvature. The boundary conditions are

$u = w = 0, \quad z = 0$ (No slip boundary condition),

$\frac{\partial u}{\partial z} = 0, \quad z = h(r)$ (stress free surface),

For constant jet flow rate *Q* the radial velocity satisfies

$$2\pi r \int_0^h u \, dz = Q. \quad (3)$$

We consider flow in the film balanced by frictional drag so that $\frac{du_s}{dr} \sim \frac{-\nu}{h^2}$, and where surface tension is dominant so that the Weber number $\sim \frac{\rho u_s^2 h}{\gamma}$, based on the film thickness, is of order one. This implies, using continuity (3), that the jump radius scales as

$$R_o = \frac{Q^{3/4} \rho^{1/4}}{(\nu\gamma)^{1/4}} \quad (4)$$

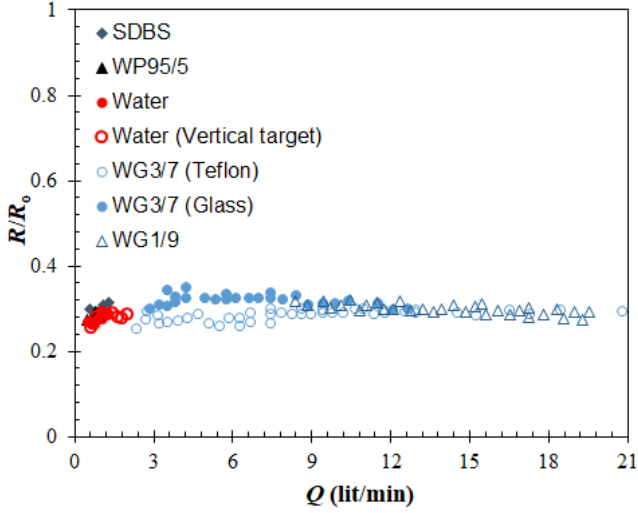

Figure 2 Dimensionless jump radius plotted against the flow rate for all our experiments with different liquids and surface orientation.

Figure 2 plots the scaled radius $\frac{R}{R_0}$ (4) against $Q$. The data from a broad range of experiments with different $Q$, physical properties and surface orientation (Table 1) all collapse onto the line $\frac{R}{R_0} \approx 0.289 \pm 0.015$. This collapse of the data implies that the dominant balance in the formation of thin-film jumps is associated with surface tension and that gravity is irrelevant.

### 3. Theory

In order to evaluate the jump condition more precisely we employ an ansatz for the velocity based on Watson's similarity profile. The radial velocity is represented as $u = u_s f(\eta)$, where $\eta = z/h$ ($0 \le \eta \le 1$) and $u_s$ is the velocity at the free surface. Hence (see Supplementary Information T1) $u_s r h =$ constant. From (1) we get

$$w = u h' \eta = u_s h' \eta f(\eta) \quad (5)$$

Writing (1) in the form

$$\frac{1}{2}\rho(u^2+w^2)\left(\frac{\partial(ru)}{\partial r}+\frac{\partial(rw)}{\partial z}\right)=0 \quad (6)$$

allows the mechanical energy equation to be written as

$$ur\frac{\partial}{\partial r}\left(\frac{1}{2}\rho(u^2+w^2)\right)+wr\frac{\partial}{\partial z}\left(\frac{1}{2}\rho(u^2+w^2)\right)$$
$$=-ur\frac{\partial p}{\partial r}-\rho g u r\frac{dh}{dr}+r u \mu\left(\frac{\partial^2 u}{\partial z^2}\right) \quad (7)$$

Integrating (7) from the bottom to the free surface, and adding, as the last term below, the surface energy term at the free surface, the RHS of (7) can be written as

$$-\int_0^h ur\frac{dp}{dr}dz - \int_0^h \rho g u r \frac{dh}{dr}dz + \int_0^h \mu u r\left(\frac{\partial^2 u}{\partial z^2}\right)dz + \gamma \frac{d(ru_s)\int_0^1 f(\eta)d\eta}{dr}$$

where $\gamma$ is the surface tension of the liquid. Inserting Watson's ansatz yields

$$\left(\rho u_s^2 h r\int_0^1 f^3(\eta)\left(1+(\eta h')^2\right)d\eta - \rho g h^2 r\int_0^1 f(\eta)d\eta - \gamma r\int_0^1 f(\eta)d\eta\right)\frac{du_s}{dr}$$
$$= -u_s r h \frac{dp}{dr}\int_0^1 f(\eta)d\eta + \rho g h^2 u_s \int_0^1 f(\eta)d\eta$$
$$+ \gamma u_s \int_0^1 f(\eta)d\eta + \frac{\mu r u_s^2}{h}\int_0^1 f(\eta) f''(n)d\eta \quad (8)$$

and the radial dependence of the surface velocity is

$$\frac{du_s}{dr}$$
$$= \frac{-u_s r h \frac{dp}{dr}\int_0^1 f(\eta)d\eta + \rho g h^2 u_s \int_0^1 f(\eta)d\eta + \gamma u_s \int_0^1 f(\eta)d\eta + \frac{\mu r u_s^2}{h}\int_0^1 f(\eta) f''(n)d\eta}{\left(1-\frac{1}{We}-\frac{1}{Fr^2}\right)\rho u_s^2 h r\int_0^1 f^3(\eta)\left(1+(\eta h')^2\right)d\eta} \quad (9)$$

Here the Weber number $We$ (comparing inertia and surface tension) and the Froude number $Fr$ (comparing inertia and gravity) are defined as

$$We = \frac{\rho u_s^2 h \int_0^1 f^3(\eta)\left(1+(\eta h')^2\right)d\eta}{\gamma \int_0^1 f(\eta)d\eta} \quad \text{and}$$

$$Fr^2 = \frac{u_s^2 \int_0^1 f^3(\eta)\left(1+(\eta h')^2\right)d\eta}{g h \int_0^1 f(\eta)d\eta}$$

Equation (9) was solved for $u_s$ with the initial condition obtained from Watson's (1964) analysis of the growth of the boundary layer. The boundary layer first occupies the whole film at $r_{bl}$, given by $r_{bl}/d_o = 0.1833 Re^{1/3}$, where $d_o$ is the nozzle diameter and the jet Reynolds number $Re = 4Q/\pi \nu d_o$. At this location $u_s$ is set equal to the mean jet velocity, and (9) provides its subsequent radial values. At the location where $We^{-1} + Fr^{-2} = 1$, (9) becomes singular and there is a discontinuity in the film velocity and the liquid film thickness changes abruptly. Therefore, the condition for hydraulic jump is

$$We^{-1} + Fr^{-2} = 1. \qquad (10)$$

This condition provides a more precise estimate than the scaling argument (4) and also includes the effect of gravity. There are two limiting cases.

Case 1: $We \approx 1$ and $Fr \gg$ . The jump is caused by surface tension, and occurs when the film thickness is small and the momentum per unit width is of the order of the surface tension. For circular hydraulic jumps, the expanding flow field favours this case and most jumps are induced by surface tension.

Case 2: $Fr \approx 1$ and $We \gg$ . When the liquid film thickness is large and the flow of momentum per unit width is high compared to the surface tension, then the jump is initiated by gravity. None of the experiments reported in this paper correspond to this case.

## 4. Experiments:

Circular hydraulic jumps were produced by liquid jets impinging normally onto a planar solid boundary. Both a vertical jet impinging on a horizontal plate from above and below and a horizontal jet impinging on a vertical wall were studied. The jet nozzle diameter was 2 mm and $Q$ varied from 0.49 to 2 l/min. For low flow rates ($Q$ <1.3 l/min), liquid was supplied from a constant-head apparatus to glass Pasteur pipettes. For higher $Q$ a centrifugal pump and a brass nozzle was used[7]. Target plates were smooth Perspex™ sheets. The vertical jet impacted a 0.25 m diameter circular disk; horizontal jets a 1.00×0.40 m rectangular plate. A Photron Fastcam SA3 was used to acquire images at up to 2000 frames per second of the liquid film and the hydraulic jump. These were subsequently processed using a Matlab™ script and ImageJ.

The viscosity and surface tension was varied by using a range of water mixtures (Table 1). The surface tension was varied by about a factor of three by using water/1-propanol mixtures (5 w/w% labelled WP95/5) and a solution of sodium dodecyl benzene sulfonate (SDBS).

Table 1 Properties of the liquids used

| Liquid label | $T$ (°C) | $\gamma$ (mN/m) | $\nu$ (cS) | $\rho$ (kg/m³) |
|---|---|---|---|---|
| Water | 20 | 72 | 1.002 | 1000 |
| WP95/5[14] | 20 | 42.5 | 1.274 | 989 |
| SDBS[15] | 20 | 38 | 1.00 | 1000 |
| WG3/7[10] | 19 | 67 | 20.7 | 1160 |
| WG1/9[10] | 28 | 65 | 99.3 | 1240 |

With incorporation of experimental data from Jameson *et al.* (2010), the viscosity and flow rate were varied by more than factors of 100 and 10, respectively. More than 120 experiments were conducted.

## 5. Results

Figure 3(*a*) compares experimental measurements with theoretical predictions of $R$ for water, WP95/5 and the aqueous SDBS solution. The SDBS and water differ in their surface tension, while the WP95/5 and SDBS have different viscosities but similar surface tensions. Lowering the surface tension (SDBS *cf.* water) increases $R$ while increasing the viscosity (WP95/5 *cf.* SDBS) reduces $R$. The corresponding theoretical curves obtained from (9) are shown in figure 3(*a*) and agree with the experimental measurements.

We studied the effect of gravity by changing the orientation of the surface. There is a small influence of gravity as evident in the non-circularity of the jump in figure 3(*b*) (inset image), however it is not significant as can be seen from measurements of the jump radius in the direction perpendicular to gravity (figures 3 (*a*) and (*b*)). The data and predictions show excellent agreement. Figure 3 (*b*) also compares the jump radius when water jets impinged from under the surface. In this case, the jump radii are slightly larger compared to the jump radii on vertical plate.

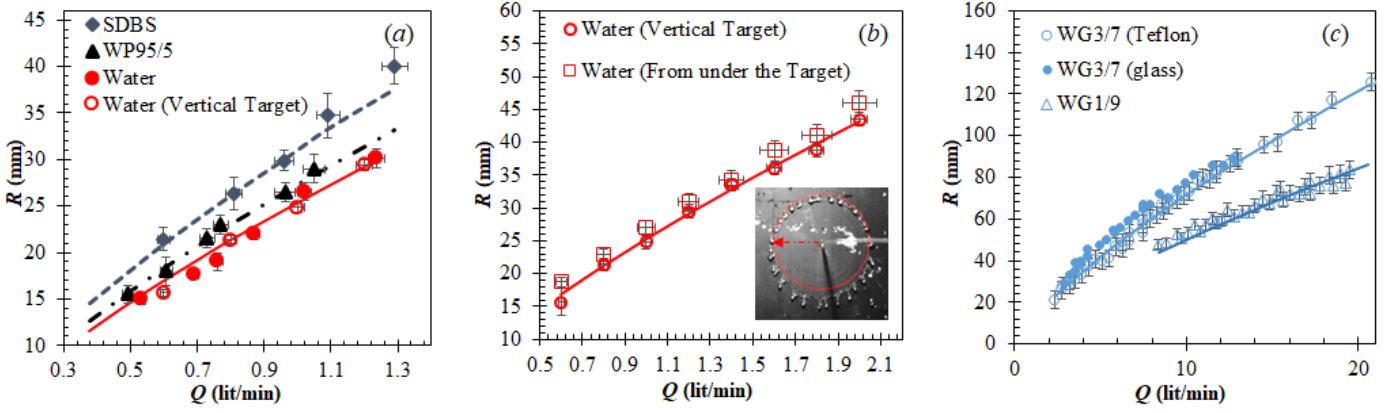

Figure 3 Comparison of the theoretical predictions (lines), obtained from solutions of (9) with the data (markers). (*a*) Location of the initial jump for normal impingement on a horizontal plate from above. (*b*) Jump on vertical and horizontal surface: for the vertical surface the radius was measured in the direction perpendicular to gravity, see inset, on the horizontal surface the liquid jet impinged from below the surface. (*c*) Water bell radius for 70% and 90% glycerol/water solutions (from Jameson *et al.* (2010))

Figure 3(*c*) compares data reported by Button *et al.* (2010) when the liquid jets using 70% and 90% glycerol/water solutions at 19°C and 28°C, respectively, impinged on the underside of the surface. For a given flow rate, the departure radius is smaller for the latter, more viscous, solution: the surface tensions are comparable. The experimental data and the theoretical prediction are again in excellent agreement.

### 6. Planar hydraulic jump

A closely related flow is the planar hydraulic jump in a thin liquid film (fig. 4). The existing theory [1] again argues that the jump occurs due to gravity near the location where $Fr = 1$. However, Leinhard *et al.* (1993) reported that the planar hydraulic jumps induced in thin liquid films are influenced by the surface tension of the liquid [13].

A theoretical explanation of these observations has not been reported, and we demonstrate that our approach for the circular hydraulic jump can also explain the formation of the planar variety.

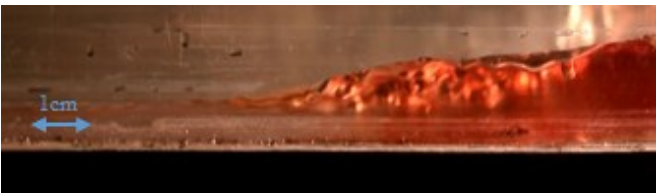

Figure 4. Planar hydraulic jump for water of initial film thickness 0.85 mm.

As for the circular case a scaling relationship for a planar hydraulic jump, matches the dissipation and surface tension energy. In contrast, however, the liquid film does not expand and even for a submillimetre thick planar liquid film, the contributions from both the Froude or Weber numbers are relevant. For a flowrate $q$ per unit width, considering the Weber number, we get $X_{o,1} = \dfrac{q^3 \rho}{\nu \gamma}$; considering the Froude number, we obtain $X_{o,2} = \dfrac{q^{5/3}}{\nu g^{1/3}}$, where $X_{o,1}, X_{o,2}$ are scaled jump locations.

As before also applying a similar analysis to this planar flow yields (Supplementary Information T2)

$$\dfrac{du_s}{dx} = \dfrac{-u_s h \dfrac{dp}{dx} \int_0^1 f(\eta)\,d\eta + \dfrac{\mu u_s^2}{h} \int_0^1 f(\eta) f''(\eta)\,d\eta}{\left(1 - \dfrac{1}{We} - \dfrac{1}{Fr^2}\right) \rho u_s^2 h \int_0^1 f^3(\eta)\left(1 + (\eta h')^2\right) d\eta} \quad (11)$$

It is clear from (11) that the initial hydraulic jump will again occur where $We^{-1} + Fr^{-2} = 1$

The initial condition to solve (11) was obtained using Blasius flat plate boundary layer equation for $\delta \approx 5\sqrt{\dfrac{\nu x}{u_o}}$. The boundary layer first occupies the whole film at $x_o = \dfrac{0.106 q^2}{\nu u_o}$.

At this location $u_s$ is set equal to the mean jet velocity, $u_s = u_o$. Where $u_o$ and $q$ are mean initial velocity and flow rate/width. The planar jump experiments employed a Perspex™ flow channel of width 0.15 m and length 2.5 m equipped with a 0.30 m high reservoir with an adjustable gate. The other end of the channel was open and liquid was discharged into a storage tank.

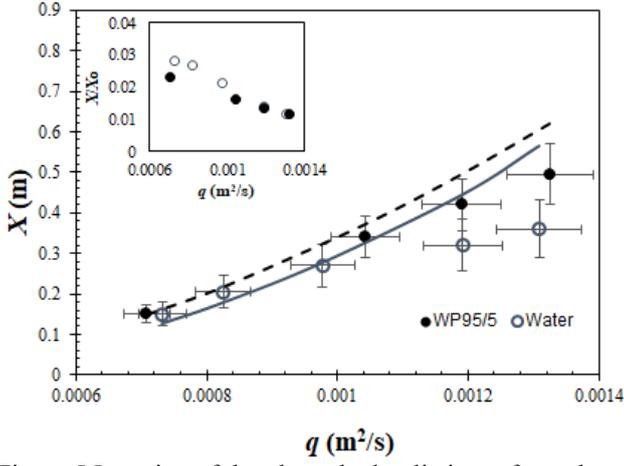

Figure 5 Location of the planar hydraulic jump for a channel of breadth 0.15 m and slit-width or initial film thickness of 0.85 mm. The inset figure shows a comparison of scaled jump location $\frac{X}{X_{o,1}} = \frac{X\nu\gamma}{q^3\rho}$ with respect to $q$.

Figure 5 compares measurements of the initial jump location with the theoretical prediction for two different liquids, a 5% 1-propanol solution and water at different $q$, (flow rate per unit width). Unlike the circular hydraulic jump, the initial planar jump does not exhibit a sharp transition and the data are more prone to experimental uncertainty, where small waves appear at the jump location which amplify and show a sharp transition. The data for the 5% 1-propanol solution nevertheless show good agreement with the prediction. For water, at higher $q$, the model over-predicts the experimental measurement. This was due to the experimental limitations with the apparatus, which prevented the flow attaining a steady value.

The inset figure shows the scaled jump location $\frac{X}{X_{o,1}} = \frac{X\nu\gamma}{q^3\rho}$, considering surface tension, with respect to $q$. In contrast to circular jumps, the scaled locations do not show a constant value. This is due to the fact that the contribution from Froude number cannot be ignored. Nevertheless, the data for water and WP95/5 show the same trend. The scaled jump location considering gravity $\frac{X}{X_{o,2}}$, show the same trend but do not collapse on same curve (data not shown).

**Conclusions**

This paper provides a theoretical resolution to the question: what initiates a hydraulic jump in a thin liquid film? For a circular jump the scaling relationship $R \sim \frac{Q^{3/4}\rho^{1/4}}{(\nu\gamma)^{1/4}}$ clearly shows that the jumps are caused by the viscosity and the surface tension of the liquid. The detailed analysis shows that the hydraulic jump, or the supercritical to subcritical transition, occurs when $\frac{1}{We} + \frac{1}{Fr^2} = 1$. From (8) we infer that the transport of surface energy becomes dominant for the expanding films at larger radii. The LHS of (8),

$$\rho u_s^2 hr \int_0^1 f^3(\eta)\left(1+(\eta h')^2\right)d\eta - \rho g h^2 r \int_0^1 f(\eta)d\eta - \gamma r \int_0^1 f(\eta)d\eta \frac{du_s}{dr}$$

indicates that the liquid momentum has to overcome the hydrostatic pressure and surface tension. The jump is formed where the hydrostatic pressure term $\rho g h^2 r \int_0^1 f(\eta)d\eta$ and surface force $\gamma r \int_0^1 f(\eta)d\eta$ are greater than or equal to the momentum. This behaviour was previously attributed to the hydrostatic force *alone*, which is a special case of the general solution.

Previous analyses have incorporated surface tension but only through the hoop stress, which, we agree, is small and unimportant, although it is included in our analysis. It is the inclusion of the loss of energy associated with the radial transport of surface energy that, due to viscous forces, implies that the flow can no longer provide the kinetic energy to maintain the thin film. At this point the flow decelerates rapidly, the depth of the flow increases and the hydraulic jump occurs. This is equivalent to the surface tension force associated with curvature of a film of thickness $h$, and hence this thickness is the relevant length scale in the Weber number used to obtain the scaling relation (4).

The critical Weber number based on the film thickness implies that the flow speed is $\sim \sqrt{\frac{\gamma}{\rho h}}$, which is the same as the speed of capillary waves $C \sim \sqrt{\frac{\gamma k}{\rho}}$, with wavenumbers comparable to the inverse of film thickness. Consequently, capillary waves play a similar role in this situation to gravity waves in the traditional hydraulic jump.

For a planar hydraulic jump, the liquid films are relatively thick with a larger Weber number, giving the jump condition as $Fr \approx 1$. For a circular hydraulic jump, $Fr$ is large and jumps are induced by surface tension. The analysis also explains the observation that the radius of the water bell departure and hydraulic jump on a vertical surface is not sensitive to the orientation of the surface. The analysis highlights the importance of relatively weak surface forces on thin liquid film such as those found in coating, cleaning and heat transfer.

**Acknowledgements**

Funding for RKB from the Commonwealth Scholarship Commission is gratefully acknowledged, as are helpful discussion with Prof J. F. Davidson. NKJ acknowledges the support of an EPSRC IAA grant. The authors also wish to thank Dr C. Gladstone for providing the flow channel and Mr L. Pratt for help in modifying the apparatus.